
\magnification=1200
\hsize=13cm
\def\rf{\parshape=2 0pt \hsize 0.7cm 12.3cm}
\raggedbottom

\def\pound{{\it \$}}
\def\<{{<}} \def\>{{>}}

\headline={\hfil}
\footline={\hss \tenrm \folio \hss}

{\bf
\centerline{On Many-Minds Interpretations of Quantum Theory}
\medskip

\centerline{Matthew J. Donald}
\medskip

\centerline{The Cavendish Laboratory,  Madingley Road,}

\centerline{Cambridge CB3 0HE,  Great Britain}
\smallskip

\centerline{ e-mail: mjd 1014 @ phy.cam.ac.uk \  \footnote*
{\tenrm quant-ph/9703008 revised November 1997.}}}
\medskip

\noindent{\bf abstract} \quad  This paper is a response to some recent
discussions of many-minds interpretations in the philosophical
literature.  After an introduction to the many-minds idea, the
complexity of quantum states for macroscopic objects is stressed.  Then
it is proposed that a characterization of the physical structure of
observers is a proper goal for physical theory.  It is argued that an
observer cannot be defined merely by the instantaneous structure of a
brain, but that the history of the brain's functioning must also be taken
into account.  Next the nature of probability in many-minds
interpretations is discussed and it is suggested that only discrete
probability models are needed.  The paper concludes with brief
comments on issues of actuality and identity over time.
\medskip

\medskip
\centerline{\bf Introduction}
\medskip

The purpose of this paper is to discuss some philosophical issues which
have arisen in ``many-minds'' interpretations of quantum theory.  The
paper begins with a brief introduction to the many-minds idea laying
particular emphasis on the complexity of quantum states for
macroscopic objects.  A more thorough introduction at an elementary
level is given in Lockwood 1996a.   The relationship between mind and
brain is of fundamental importance to many-minds interpretations and in
the second section, I shall argue that, in the light of quantum theory, we
should revise our understanding of what a brain is.  In particular, I shall
suggest that the history of a brain's functioning is an essential part of
its nature as an object on which a mind supervenes.   Then I shall turn to
the issue of probability in many-minds interpretations and argue against
the claim that there should be a continuous infinity of minds at each
instant.  I shall also make a number of briefer comments on other
philosophical questions.  In three long and technical papers (Donald
1990, 1992, 1995) I have presented a version of the many-minds
interpretation which I believe to be compatible with special relativity,
with quantum field theory, and with the macroscopic and thermal nature
of real observers.  The ideas discussed here stem from that technical
work but also stand independent of it.

	Many-minds interpretations are a class of ``no collapse'' interpretations
of quantum mechanics, which is considered to be a universal theory.
This means that they assert that all physical entities are governed by
some version of quantum theory, and that the physical dynamics of any
closed system (in particular, the entire universe) is governed entirely by
some version, or generalization, of the Schr\"odinger equation.  From the
point of view of a committed quantum theorist, who knows neither of
any experimental evidence for any breakdown in quantum theory, nor
of any alternative theory which is not both ad hoc and incompatible
with special relativity, these assertions may seem plausible
(cf.~Deutsch 1996).  It is certainly the case that, at least over short
time intervals, quantum states can be found which will give apparently
accurate representations of the physical states of essentially all
non-gravitational physical systems, however large or complex they may
be.  Indeed, states can be found which are not only ``apparently accurate''
in the sense that they are compatible with all our actual short-term
observations, but also in the sense that they represent any particle as
being, during the interval considered, sufficiently well-localized to
accord with conventional pictures of that type of quantum particle.
Thus, in such states, the atoms in our environment are represented as
slightly fuzzy balls, while free electrons are represented as moving
along slightly fuzzy straight lines.  I shall call these states
``pragmatic''.  For example, a pragmatic state for a gas might use
minimum uncertainty wave-packets centred on definite choices of
positions and momenta to describe the centre of mass variables of the
molecules, together with appropriate choices of molecular
wavefunctions for the electronic variables.  At a more sophisticated
level, using our excellent understanding of quantum states for single
molecules, and the possibility of building such quantum states up to
describe many molecules, pragmatic quantum states can be ascribed at
any time to any chemical system -- including the human brain.   The
fundamental problem -- the problem of ``Schr\"odinger's cat'' -- arises
because the time continuation given by the quantum mechanical dynamics
does not always lead from a quantum state which provides an apparently
accurate description at one time to a quantum state which provides a
similarly apparently accurate description at later times.  The
pragmatism of a state, in other words, is time dependent.

	Consider, for example, electrons contributing to the production of an
interference pattern by passing, one at a time, through some kind of
two-slit device and hitting a position detector.   (Pictures from such an
experiment are presented in Tonomura et al.~1989.)  I have chosen this
example, partly because the pictures are such a direct demonstration of
the difference between quantum state and observation, partly because
the electrons are more likely to be seen to hit some parts of the detector
than others, and partly because I think that the two-alternative
experiments by which this subject is usually introduced foster a na\"\i
ve view of the complexity of quantum states (cf.~Weinstein 1996).

	According to quantum mechanics (by which I shall henceforth mean
universal, no collapse, quantum mechanics), nothing determines the
points on the detector where the electrons hit.  Indeed, if a pragmatic
state is given for an electron at the beginning of the experiment, then
the slits act to amplify the fuzziness of its original trajectory, so that
as it approaches the detector, the state associated with it will
represent some sort of weighted distribution over all possible hitting
points.  The ``many worlds'' idea is that such ``superpositions'' of
possibilities can form part of a correct and complete description of the
physical world and that all possible hittings do, in some sense,
happen.  This is an interesting idea, but it leads immediately to
questions as to why each electron is only ever seen to hit at one point
and to what the probability of any particular hitting being seen means.
That the word ``seen'' slips naturally into these questions, suggests that
here, as in so many other attempts to understand quantum mechanics,
``the observer'' has some special role to play.  Everett himself said (in
DeWitt and Graham 1973, p.~117) that, within the context of his theory,
``it develops that the probabilistic aspects of [von Neumann's] Process 1''
(the collapse of the wavefunction) ``reappear at the subjective level, as
relative phenomena to observers.''

Consider, therefore, someone who is looking an image produced directly
by the detector, or even someone who is looking at one of the
photographs reproduced from Tonomura et al.~1989 on page 3 of
Silverman's 1995 book.  They will see a picture of a definite pattern of
specks.  Nevertheless, if we consider the quantum states that we should
associate to the physical structures of these human beings, and, if we
start by looking back in time as far as the beginning of the experiment
and take as our initial condition a ``pragmatic'' state for the entire
situation at that time, then, solving the Schr\"odinger equation will lead
to quantum states which by the time of the observations considered will
also describe some sort of weighted distribution; in this case over brains
seeing possible patterns of specks.  A many-minds theory aims to accept
such ``unpragmatic'' states, and to interpret them.

The first step in a many-minds interpretation is to make the hypothesis
that our nature is such that we cannot see such a picture except as a
definite pattern.  The quantum state is correct in its description of
a weighted distribution of possible patterns.  When a single individual
comes into contact with the picture for the first time, all those
different possibilities do occur, but each different pattern is seen by a
different mind; minds which share the same past and the same name, but
which experience different presents and different futures and which
have no means of communicating to each other.  The probability of seeing
a given pattern is determined, at least to a first approximation, by the
corresponding weight in the quantum state.

	In any no collapse interpretation, including the modal interpretation and
the Bohm interpretation as well as all versions of the many-worlds
interpretation, we start off with only one quantum state:  the uncollapsing
universal quantum state, which I shall denote by $\omega$.  $\omega$
can be identified by going backwards in time.  Each time we pass back
through the appearance of a collapse we get a better approximation to
$\omega$.   Eventually, we arrive back at the big bang.  For the moment,
we may ignore the question of whether the big bang itself was merely the
appearance of a collapse.  The quantum state of the universe coming out
of the big bang looks -- at least in its non-gravitational aspects -- very
like a thermal equilibrium state.  In the Hamiltonian (uncollapsed) time
propagation of that state, the stars and planets which we see now do not
exist as definite objects, and certainly neither does any particular
measuring device now being used by us on one of those planets.
$\omega$ seems to be a complete mess.  However, it does have a great
deal of hidden structure, and it is the job of a no collapse interpretation to
explain how that hidden structure comes to be seen.

Most workers in no collapse interpretations have produced no more than
elementary models based on the definite existence of specific measuring
devices.  They have assumed, for example, that the Hilbert space of the
universe splits naturally into a tensor product structure compatible with
the measurement under consideration.  They have also assumed,  even
when describing the behaviour of macroscopic objects, that it is
appropriate to employ models in which only a few dimensions of Hilbert
space are used to describe all the relevant behaviour.  In my opinion,
these assumptions are untenable (cf.~Bacciagaluppi, Donald, and Vermaas
1995).  The first assumption begs the question of what is natural, and
depends on assumptions about the nature of particles which are known to
be false according to relativistic quantum field theory (Haag 1992).  As
far as the second is concerned, a measure of the number of dimensions
relevant to a macroscopic object can be given by
$e^{S/k}$ where $S$ is the absolute thermodynamic entropy of the object
and $k$ is Boltzmann's constant.  This number is so large that it must call
into question any argument which refers to ``the wavefunction
($\psi_{observer}$) of the observer''.  Possible wavefunctions do exist,
but there are something like $e^{S/k}$ orthogonal choices available, and
any one of these choices will, by entanglement with the environment,
rapidly move into a mixture with the other choices.  This vast complexity
of available observer states suggests, for example, that Weinstein
1996 is quite right to criticize Albert and Loewer's 1988 invocation of
a projection measuring a belief held by a observer.

The analysis of $\omega$ is not however a hopeless task.  The secret lies
in the idea of ``correlation''.  Quantum states, particularly many-particle
quantum states, are patterns of correlations.  For example, consider  the
quantum state describing a volume of hydrogen and oxygen atoms in a
two-to-one mixture which has come to chemical and thermal equilibrium
at room temperature and pressure.  The state is a canonical ensemble
equilibrium density matrix.  It is not a pure state, but that is only to be
expected, because states of non-isolated macroscopic systems are
virtually never pure.  (As I discuss elsewhere, this lack of purity
does not solve the conceptual problems we are considering (Donald 1990,
1992).)  The important point is that it is not a pragmatic state because
it does not describe a collection of water molecules with well-defined
positions.  Indeed, the position distribution of each individual atom
considered separately will be spread uniformly over the entire
container.  Nevertheless, if the position of one of the oxygen atoms is
taken as given, then the same equilibrium quantum state determines a
distribution for the other atoms which will be such that precisely two
hydrogen atoms will be closest to the given oxygen position, and those
two will be around 0.096nm distant.  Similarly in the Schr\"odinger cat
state -- if there has been a radioactive decay, then the cat is dead; if
there is broken glass, then the flask which contained the poison has been
broken, and the cat is dead; if, as he richly deserves, the experimenter's
face is scratched when he opens the steel box, then the radioactive
atoms have not yet decayed.  An adequate analysis of the correlations in
$\omega$ is the first step towards an interpretation of quantum
mechanics.

\bigskip
\centerline{\bf The object of supervenience}
\medskip

	If we are to make sense of the many worlds idea and demonstrate how
the world that we see is extracted from $\omega$, it would appear that
we need to be able to say what a world is.  Saunders 1996b has pointed
out that there are three ways of approaching this problem.  One is direct
attack, by mathematical definition.  In my opinion, ``consistent histories''
is an example of this approach.  So far, however, I do not believe that this
direct approach has been successful.  Dowker and Kent 1996, in
particular, have found many problems with the consistent histories
program.  It is not clear that the abstract definition of a consistent
history does solve the problem of defining a world, because there are
many possible sets of consistent histories, and no way of choosing one
particular set is available.  The second approach is the ``many minds''
idea.  Here we give observers a central role.  We do not need a general
definition of ``worlds'', but we do need to define ``observers''.  The only
worlds which are considered important are identified by correlations to
those observers and, as Brown 1996 has emphasized, these external
worlds are secondary, derived, concepts.  The final approach is the idea,
championed by Saunders 1995, 1996a,b, of ``relativism''.   The claim
here is that we do not need to define ``worlds'';  the total pattern of
correlations established by the universal quantum state are sufficient in
themselves as a foundation for physics.  Saunders claims consistent
histories as an example of relativism.  As it stands at present, I would
accept that it is possible to interpret consistent histories in either way.

Saunders' approach might seem to avoid the problem of definition, but, in
my opinion, this is not a problem that we should be trying to avoid.
Although the correlations established by $\omega$ are enough to provide
a description of objective reality, they do not give us any understanding of
subjective reality. We are here and we look out at the universe entirely
through those particular correlations which $\omega$ establishes to our
brains.  These correlations are the ones that count.  We do not need
another ``view from nowhere'' in physics.

I also do not believe that we exist merely ``for all practical purposes''.
In a state of water at equilibrium, I can localize the oxygen atoms, to a
certain extent, by giving the positions of the hydrogen atoms, or I can
localize the hydrogen atoms, to a certain extent, by giving the positions
of the oxygen atoms.  Similarly, in Saunders' program, the state of our
brains determines for all practical purposes the state of the world that
we experience, and the state of the world that we experience determines
for all practical purposes the state of our brains.  But this raises the
as-yet-unsolved technical problem of producing a theory, at the level of
$\omega$, of the ``quasi-classical domains'' in terms of which, for all
practical purposes, our experienced world is to be described (Saunders
1993, Gell-Mann and Hartle 1993).  Not only does it seem to me that it
is more straightforward to solve the technical problem of finding an
unambiguous characterization of an observer (Donald 1995), but also I
believe that consciousness is a fundamental pillar of existence and must
therefore be something definite.

In my opinion, I am what I am, and I want to discover what that what
might be.  In particular, I want to discover exactly what the physical
manifestations of that what might be.  Which physical properties are
fixed by the existence of a given state of my mind, and which properties
are only probabilistically constrained?  Here the ontological issue that
``I am what I am'' must be distinguished from the epistemological
question of whether I can know what I am.  No conscious being can be
aware of his exact state, but this does not mean that he does not need to
be in some exact state.   I think that it is the proper task of an analysis
of quantum theory to try to give an exact definition of the possible
physical manifestations of an observer.

The doctrine of psychophysical supervenience claims that two people
cannot differ mentally without also differing physically (Lewis 1986,
p.~14).  In a classical picture, brains are just there, and the investigation
of exactly on what aspects of those brains the corresponding minds
supervene hardly seems to be an essential task.  In quantum theory,
however, the universal quantum state is a grand superposition of
possibilities.  Some of those possibilities do seem to contain structures
which look like brains or which function like brains.  Now the doctrine of
supervenience no longer merely provides a convenient name behind which
questions about the nature of mind can be hidden.  Instead, it raises the
difficult but intriguing technical problem of analysing such a
superposition (cf. Albert and Loewer 1988, Barrett 1995, Lehner
1997).  This requires discovering what the relevant physical
constitution of a person might be.  In other words, the problem is to
discover a physical ``object of supervenience'' such that the doctrine of
supervenience holds in the form that two people with identical ``objects
of supervenience'' are mentally identical.  Putting this another way, the
idea of psychophysical parallelism encounters the problem of identifying
just what it is to which the psyche is supposed to be parallel.

One way of approaching these problems is to consider the preliminary
claim that two brains which are physically sufficiently similar,
necessarily give rise to the same mental phenomena.  Even without any
analysis of ``necessarily'', ``the same'', or ``mental phenomena'', this
claim seems plausible.    For example, it seems to me to be plausible that
mental phenomena would not change if the temperature of the brain was
changed by less than one thousandth of a Celsius degree, nor if the pH
changed by less than $0.01$.  So what exactly does  ``physically
sufficiently similar'' mean?  What physical structures could be used to
define such equivalence classes of brains?  Can we describe the minimal
amount of structure required or discover a simple characterization of the
equivalence classes?  Initially, I shall refer to anything which is a brain
by virtue of having such a structure, as being  ``an object of
supervenience'', although I am working towards a position in which,
strictly speaking, the term should be reserved for the equivalence
classes rather than for the members of those classes.

	It is important that the words ``give rise'' in the claim also be taken
seriously.  There are two different techniques by which it is possible to
think about what it means for a physical situation to imply the existence
of mental phenomena.  The more conventional technique is to imagine
strange ways in which such a situation could arise and to contemplate
whether the proposed mental phenomena remain plausible.  For example,
someone who claims that it is only the instantaneous structure of a brain
which has any importance, could be required to consider the possibility
that a functioning piece of flesh could be constructed merely by bringing
the requisite atoms together, for a moment, in a vat.  A more powerful
technique, however, is to imagine the mind as a ``ghost in a machine'' and
to ask what properties of the machine the ghost would use to find its
meaning.   This is not to suggest that the ultimate aim is not to exorcize
all such ghosts.  The object of supervenience exists for itself.  However,
if the physical structure is sufficient to imply the meaning, then the
meaning must be interpretable from the physical structure.  Imagining a
ghost is merely a way of thinking about what might be required in the
interpretation of physical structure.  For example, if the doctrine of
supervenience is to have any force, then the meaning should be
interpretable without requiring that the ghost be an educated human
being with a training in twentieth century neurophysiology and
nineteenth century physics who already knows that the machine is a
functioning human brain.

	For many of those who have written on the many-minds interpretation,
the object of supervenience is an element of a basis of brain states.  For
example, Lockwood 1996a refers to a ``consciousness basis'' and Albert
and Loewer 1988 to an incomplete basis of ``brain (or brain +
environment) states''.  In neither case, is the particular basis identified.
In my opinion, the attachment to the idea of a basis is a mistake based on
a false analogy with elementary models of measurement theory.  A
human brain is not like an atom in a Stern-Gerlach device, nor is it like a
Copenhagen-interpretation measuring device which bears on its cover the
name of the self-adjoint operator which it is designed to measure.  A
brain is warm and wet.  The number $e^{S/k}$ of available
wavefunctions is something like $10^{10^{26}}$.  The identification of a
suitable basis might be expected to require identification of exactly which
molecules are to be included in any given basis state.  Is this set of
molecules to include the blood which is pumping through the brain, and
the molecules which it is breathing?  Where is the outside surface of the
brain (or in Albert and Loewer's case of the brain + environment) to be
drawn?  Is Albert part of Loewer's environment, and if not, where is the
line of separation?  Why should any, arbitrarily small, change in pH --
which changes the number of hydrogen ions in the brain, and therefore
changes the basis state -- be sufficient to change the object of
supervenience?

	Many-minds authors have also tended to assume that the object of
supervenience is a brain at an instant.  This too is an assumption with
which I disagree, and I continue to disagree, even if ``the instant'' is
modified, as Butterfield 1996 proposes, to cover the duration of a
psychological moment.  The doctrine of supervenience has become
dominant in modern philosophy, at least in part because of the success of
neural computation models of the brain (e.g.~Dennett 1991, Churchland
and Sejnowski 1992).  In the framework of classical physics, these
models are part of an increasingly convincing demonstration that
everything that a mind appears to understand or feel is reflected in detail
in neural functioning.  However, although behaviour may be governed
entirely by instantaneous neural functioning, this does not imply that
instantaneous functioning is sufficient for  mental understanding.
Indeed, in my opinion, even short-term functioning carries, in itself, very
little meaning.

Imagine that you were given a perfect snapshot of a brain.  How would you
begin to understand the information that was being processed?  How
would you understand the dispositions of the individual under study?
Suppose that you find excitement in one part of the brain, for example,
the occipital lobe.  The name merely tells you that this is at the back of
the head.  Tracing nerve cells will connect this part to a part at the front
of the head which we call the retina, and which you and I know to be
involved in vision.  How would a ghost in the brain faced with such a
snapshot know what the retina is for?  How would he know in which
direction nerve cells should be traced?  How would he even know that
nerve cells  -- rather than glial cells, or blood vessels, or the positions
of individual atoms, or electrons -- are what he should be studying?

  Supervenience suggests that it be possible to make explicit all the
information that a ghost would require for understanding.  Suppose then,
that we tell the ghost roughly what a brain is and how it functions;
suppose, for example, that we give him a textbook of neurophysiology.
With a perfect snapshot of the brain, he could then perhaps make a model
which he could try to use to find out how the actual brain under
consideration was functioning.  He would have to run that model under all
sorts of different, but physiologically ``normal'', conditions in order to find
out the details of how it worked and exactly what it was doing and
remembering at the instant of the snapshot.   The required knowledge of
the conditions of operation of the brain and the results of his simulations
would already seem to go far beyond the physical instant.  Exactly the
same types of simulation would also be required to explicate what a brain
was doing given a ``psychological moment'', or even to understand
objectively how long such a moment should last.

It would only be possible not to go beyond the physical instant, if human
brains were ``off-the-shelf'' devices for which one could, in theory,
provide a handbook mapping instantaneous molecular structure into
function.  However, human brains seem to be more like ``neural net''
machines, whose function is best discovered by simulation, than they are
like progammable computers, whose function can be explained more
compactly by provision of the manual and the program.  This applies in
particular to the details of the functioning.  Moreover, even considered
classically, human brains do not work deterministically at the level of
neural processing.  In the details of its functioning, a brain is metastable
and information is processed by the accumulation of small and uncertain
effects.  Thus we cannot tell by looking at a brain, how it has arrived at
its current state.  Different, physiologically normal, prior histories could
lead to physically identical brains.  It seems to me to be a bold
assumption that these different histories would necessarily result in
identical awareness of the same present.

My preferred alternative is to take the object of supervenience to be the
entire lifetime history of the brain in question up to the moment of the
snapshot.  Michael Lockwood has commented that this gives a picture of
us as dragging our histories behind us like ``Marley with his chain''
(Dickens 1843).  The relevant history of a brain is a history of patterns
of neural firings.  An explicit description of a history of patterns would
be a much simpler description of the information required by a ghost
than would be a textbook of neurophysiology and a full record of
simulations.  It would be simpler in two senses.  It would be shorter,
because a full record of simulations would need to analyse all the
possible behaviours of the brain, and not just the actual past behaviour.
It would also be more abstract and less contingent:  a history of
patterns, for example, would not even require a carbon-based
biochemistry for the object of supervenience.  Such a history may be
thought of as a minimal structure sufficient to define the causal
relations required by functionalism.

An appeal to the idea of functionalism seems to me to be an important
part of the hypothesis that the brain is, in some sense, an adequate model of
the mind.  (For more about this hypothesis in the present context, see the
non-technical second section of Donald 1995.)  Most many-minds authors,
although paying lip-service to materialism, seem to me to lose sight of
this hypothesis by referring to instananeous mental states distinguished
by unexplicated labels (e.g. Albert and Loewer 1988 ``belief states'',
Lockwood 1989 ``phenomenal perspectives'', Page 1996 ``perceptions'').
Barrett has advised me that ``belief states'' should be interpreted as
``dispositions'' (Albert 1992, p.~129), but once again I think that our
``dispositions'' can be more parsimoniously represented by an account of
what we have done, or said, or thought, than by an account of what we
might do, or say, or think.  The theory I am proposing is not functionalism
because I require that the causal relations be incarnated according to
specific quantum theoretical rules -- rules explained in the technical
sections of Donald 1990 and 1995.  It is also more explicit than
conventional functionalism in that the mental is taken to be constructed
from a specific type of finite pattern of causal relations between
elementary events.  My dream is that such an explicit and finite
formulation could ultimately be used to reduce functionalism to a kind of
linguistics concerned with the study of possible meanings of finite
stuctures built up from elementary ideas like ``this is the same as that'',
``this is not the same as that'', ``this is that continued indefinitely'',
``this is pleasant'', ``this is not pleasant''.  Using this linguistics, I would
hope that it would possible to argue that any ``instantaneous'' mental
state could be interpreted as the culmination of something like a (very
long) book written in a language of such elementary ideas.  However
while the books we read have only a one-dimensional causal structure --
one word comes after another -- the structures I have proposed would
also allow for  spacelike separated ``words''.

In as far as we are ghosts stuck in our brains, I think that we do make
sense of our present pattern of neural firings as a development of earlier
patterns.  It seems to me to be wrong to suggest that one particular
pattern is a tasting of a rather too strong cup of coffee, merely because
of the present arrangement of the atoms in my brain.  I have become
aware of the meaning of arrangements of atoms in my brain, not by
a process of analysis, but by existing as that functioning brain,
throughout a lifetime of many cups of coffee.  It will perhaps be objected
that I cannot, in fact, remember anything which is not somehow
contained in the present arrangement of atoms.  I accept that.  However,
the question is not what I can remember, but how I give meaning to my
physical structure.  I do not activate memories of previous cups of
coffee in order to discover what I am now tasting.  Instead, by the
activation of patterns in my brain which are correlated with earlier
patterns (and hence with earlier drinks), I become a tasting of coffee.
Consciousness develops.  It is not born anew at each instant.

	It would not matter greatly to our understanding of classical physics
whether the object of supervenience was a brain at an instant or the
history of a brain.  However, in the many-minds interpretation, the
distinction is absolutely fundamental.  A history of a brain cannot be
recorded in a single wavefunction.  As has already been discussed,
although it is always possible to find wavefunctions which form
apparently accurate representations of a brain as it can be observed, or
experienced, at one instant, there are no single wavefunctions which are
simultaneously apparently accurate representations of the brain as it is
observed over periods long enough to include, for example, performance
of the electron interference experiment mentioned above.  Indeed, in
Donald 1990, I argue that ``pragmatic'' states in the brain would have to
be replaced on a very short (e.g.~millisecond) time scale, whether or not
the brain was involved in the observation of quantum experiments.  Thus
if the object of supervenience is to be a history, then it has to be
represented either by some sort of sequence of quantum states, or, as in
consistent histories, by a sequence of projections.   (In fact, these
alternatives are, to some extent equivalent because of the duality
between states and operators.  For example, in my own technical work, in
which an explicit construction of a quantum model of the history of a
brain is given, the sequences of quantum states invoked are themselves
defined by corresponding sequences of projections.)

	I suggested above that the proper task of an analysis of quantum theory
is to try to give an exact definition of the possible physical manifestations
of an observer.  Ultimately, I think that any many-minds program aims to
characterize an observer abstractly as an information-processing
structure and to explain how that structure manifests itself physically
as some sort of objectively real quantum mechanical structure
probabilistically constrained by some sort of universal quantum state
$\omega$.  For example, an observer might be manifested by a physical
system with a wavefunction which was an element of  a ``consciousness
basis'', and abstractly characterized by a definition of such a basis.  In
my own work, in order to deal with the wide range of imperceptible
variations in possible descriptions of the physical structure of a given
observer, I have found myself moving ever further from the ``pragmatic
wavefunction'' picture of elementary quantum mechanics.  For example,  I
have moved from sequences of quantum states, corresponding to a brain
history, to abstractly characterized sets of sequences of quantum
states.  This is a progression away from na\"\i ve physical realism
towards a position in which the the physical world external to the
observer exists only as something which provides (observer-independent)
probabilities for his (objectively real) present and future existence.  The
progression is driven by the aim of finding an exact ``object of
supervenience''.  The elements of my sets of sequences of quantum states
are sequences which cannot be distinguished by the observer.  Yet each of
these indistinguishable sequences has the same relation to the universal
quantum state, and should have the same ontological status.  I am,
incorrigibly, what my mind is.  That subjective incorrigibility
corresponds objectively to something definite which is governed by
physical law.  I am not an approximation.  If I am a set of possible brain
states, then I am that set, not some element of that set.  This position
may be contrasted with that of Lockwood 1996b, p.~458, who refers to
calculating transition probabilities by considering sets of minds, which
he refers to as sets of ``identical maximal experiences''.  In as far as
there are such sets of {\sl identical} experiences, I would associate
them with the {\sl same} mind.

At the abstract level -- the level of functionalism -- I associate any
observer with a {\sl finite } structure.  Moreover, within a given bound on
complexity, only finitely many observers are possible.  I shall argue
below that this finiteness is crucial in understanding probability in a
many-minds interpretation.

  It is possible to take the progression away from na\"\i ve physical
realism further than I have previously suggested, to a point at which the
definiteness of the universal quantum state itself is called into question.  I
calculate probabilities by maximizing likelihood over indistinguishable
possibilities.  This maximization can be extended to allow variation in
$\omega$.  In this way, I believe that it may be possible to arrive at a
physics with no a priori physical constants at all.   According to such a
theory, our observations fix the value of physical constants for us, in
exactly the same way that our observations fix for us the particular
position at which an electron has hit a position detector.  If this goal can
be achieved, then it may indeed be claimed that we have a theory in
which ``arbitrariness'' or ``contingency'' has been reduced.   Lewis (1986,
\S 2.7) argues that modal realism does not reduce arbitrariness on the
grounds that the existence in our world of a specific value for a physical
constant would be as arbitrary as the unique existence of that value.
However, in the theory which I am sketching, physical constants would
not have precise values in ``our worlds''.  There might, for example,
depending on what information was available, be no fact of the matter
about whether the reciprocal of the fine structure constant was for a
given observer closer to $137.03601$ or to $137.03602$, let alone about
whether it was a rational or an irrational number.  Only the finite amount
of information which determines one's structure as observer would
determine the ``world'' in which one lives, while only a finite set of
axioms, containing no arbitrary constants, would determine the set of
possible worlds.

\bigskip
\centerline{\bf Probability}
\medskip

There has been much discussion of the meaning of probability in the
many-minds interpretation.  Here I agree with Butterfield 1996 that
specific definitions of probability measures can be justified both by
formal and by dynamical considerations.  In Donald 1992, I present
formal justifications for probabilities in a many-minds theory.  Papineau
1996 has stated that correct probabilities  ``(1) have their values
evidenced by frequencies, and (2) provide a decision-theoretic basis for
rational decisions.'' He goes on to say that these stipulations have no good
justification.  As far as the justification of (1) is concerned, many authors
have pointed out that the laws of large numbers are circular.  These laws
show only that there is negligible {\sl probability} of a long-run relative
frequency diverging significantly from the probability which is to be
justified (cf.~Kent 1990, who gives a critical survey of quantum
mechanical versions of these laws).   Here I would merely comment that,
at least in quantum theory, negligible probability is the same as small in
the topology of the space of quantum states.  This topology is a
cornerstone of the entire theory of quantum physics.  Thus, any
justification of quantum physics as a whole, including, for example, from
over-all consistency or beauty, may help to justify specific numerical
probabilities.  Although DeWitt, (in DeWitt and Graham 1973, p.~168)
was certainly wrong to claim that the ``mathematical formalism of the
quantum theory is capable of yielding its own interpretation'',
nevertheless, it seems to me that, because the topology is so natural and
fundamental a part of the theory,  arguments based on the topology can
be at least as intuitively satisfying as counting arguments in a theory of
equiprobable events.  We may not be able to explain why the world we
happen to have experienced was typical, but, if we can give meaning to
the idea of a ``world'', then at least it is possible for ``typical'' to be
well-founded and consistent.

	As far as the basis for rational decisions is concerned, the question is
whether or not, if we accept a many-minds theory, it makes any
difference to how we make decisions.  Here, despite Papineau's 1995 and
1996 arguments that it {\sl should} make no difference, I do retain a
suspicion that, if one does take many-minds theory seriously, then it does
make a difference.  I think myself that it makes one think more carefully
about {\sl all} the possible outcomes of an action.  If all futures happen,
then one cannot get away with anything any more.  A lucky escape in this
world is merely confirmation that many other worlds are unpleasant.
Perhaps if we accept many-minds then it ought to lead us to be more
``rational''; to take seriously, for example, the very large negative utilities
that we might attach to some very unlikely events.  Whether this is a good
thing or not depends on how we draw the line between being careful and
being neurotic; how we manage to accept the reality of risk without being
overwhelmed by it.

	Papineau 1995 claims that it ``is simply a basic truth about rational
choice'' that ``rational agents ought to choose those actions which will
maximize the known objective probability of desired results''.  Lockwood
1996b responds that ``choosing those actions which maximise the {\sl
expected} return, means maximising the total {\sl actual} return, as
integrated over the {\sl successors} of whatever instantaneous mind is
making the decision''.  My problem with this response is that I can see no
reason to integrate over successors.  However much an individual may
believe in a many minds theory, all he can experience, and all he is ever
going to experience, is one mind.  Suppose, for example, that you were
given the opportunity to take part in a single trial in which you would win
\pound 1000 with probability 0.6 and lose \pound 1000 with probability
0.4.  Your expected return would be \pound 200.   But, whatever may happen,
you are not going to experience receiving \pound 200.  Your only possible
experiences, whether or not your future contains both, will be that of
winning a large sum and that of losing a large sum.  Even if the monetary
quantities are accurate representations of the personal utilities to you of
the individual events of winning or losing, it is these individual
possibilities which matter and not their average.

	Suppose that we accept Papineau's view of the difficulty of justifying
probability.  It would not follow that we do not understand what it is like
to experience probabilistic events.  Even young children enjoy games of
chance and can eventually learn that their wishes do not affect the throw
of a die.  In my opinion, it would be sufficient for a many minds theory to
provide a theory of transition probabilities according to which we
experience reality as being like watching a particular, identified, stochastic
process. My own published work does not at present succeed in achieving
this goal.  Nevertheless, by modifying some technical details in my
approach to identity over time and by an analysis of the finite number of
immediate descendants of a given minimal ordered switching structure
(Donald 1995, Section 5), I now believe that it is possible to modify it
so that it does.

	Lockwood also claims that his theory can be modified to achieve such a
goal.  There is, however, a fundamental difference between the types of
stochastic process which we invoke.  My work is based on the calculation
of a priori probabilities of existence for completely and finitely specified
individual observers with finite-step histories which are completely
specified up to a given moment.  These a priori probabilities are in general
non-zero.  Thus I am proposing a theory of transition probabilities which
says that we experience the world as being like, for example, the
development of a random walk on a lattice.  We know what it is like to
watch such a process.  Examples can easily be constructed on a computer,
or even just by throwing a die.  Lockwood 1989 and Albert and Loewer
1988, on the other hand, require the existence of an uncountable number
of minds at each instant.  It is impossible to assign non-zero probabilities
at any time to more than a countable number of distinct entities.  If there
were an uncountable number of minds, then experience would be like
Brownian motion, but it seems to me that we have no idea of what it
would be like fully to experience such a process.  Brownian motion is a
continuous time process on a continuous state space.   We cannot
construct explicit physical models of continuous processes of this type,
which involve infinitely many states.  We can only make finite models.
Sample paths of a genuine Brownian motion have infinite complexity.
Indeed, almost all of the individual elements of any infinite set must have
infinite complexity in some sense.  I certainly do not know what it would
be like to watch, in all its detail, something with infinite complexity.

  This returns us to the question of Lockwood's sets of ``identical maximal
experiences''.  If it is not accepted that those sets themselves are the
objects of supervenience, then what is it like to be a mind?  Probabilities
are assigned to measurable sets.  Which measurable sets are meaningful in
Lockwood's theory? (cf.~Loewer 1996.)  Does it, for example, mean
anything that I might be in one half of a given set of  ``identical maximal
experiences'' rather than another?  What does distinguish one individual
object of supervenience from another?

	Lockwood 1996a proposes a continuous infinity of minds becauses he
argues that if there were only a finite set of minds then the probability
distribution ``should'' be uniform.  He maintains this argument in the face
of objections by Saunders 1996b, by Papineau 1996, and by Butterfield
1996 (the last, in particular, remarking that the principle of indifference
is ``a notorious dead horse'').   Lockwood 1996b appears to insist that
probability on a finite number of possibilities all of which are actual
must {\sl necessarily} be a simple matter of counting, on the grounds that
``where it is stipulated that the history of a mind, beyond a certain point,
has just $n$ discrete continuations, all of which are actual \dots there is
no freedom \dots to partition this $n$-fold continuation in any other way
than that stipulated''.  In my opinion, however,  there is no reason why
with a finite set of actual minds, the probability of one mind should not
be greater than that of another.  The a priori probability of a given mind
is part of the objective structure of the universe.  It is a number.  A
central concern of physical theory is to define such numbers.  The
numbers which arise will depend on the details of the theory, and their
justification will be a fundamental part of the justification of the theory.
In a theory which uses a consciousness basis $(\varphi_i)_{i\geq 1}$ and
a universal wavefunction $\psi$, the numbers would be defined as
$|\<\varphi_i|\psi\>|^2$, and would be justified by arguments and
evidence for square amplitudes as probabilities.  The axiomatic definition
which I give in Donald 1992 and 1995 is more sophisticated, and depends
on the past physical structure of the observer in question and on the
universal state $\omega$.  Nevertheless, ultimately, the justification of
my definitions once again stems from, or is parasitic on, the evidence for
the standard probabilistic interpretation of quantum states.

\medskip
\centerline{\bf Actuality}
\medskip

	I am very grateful to Jeremy Butterfield for the interest he has shown in
my work, and for his kind comments about it (Butterfield 1995, 1996).
In Butterfield 1996, however, he express a dissatisfaction with my
position on the issue of actuality.  Let me then address that issue here.
Butterfield claims that this is an issue ``irrespective of mind'' and which
``applies equally well to `many worlds' versions of Everettian
interpretations''.  I disagree with this claim.  In a many worlds
framework, the question is whether one should accept all possible worlds
as being actual.  I do not think that anything hangs on how one answers
this question.  In a many minds framework, however, the question is
whether one should accept all possible minds as being actual.  In this
case, there is a fundamental issue at stake.  I can see no plausibility in
solipsism.  (Why me?)  Any of your possible minds and any one of my
possible present-time minds which shares part of my past but which is
not what I am now experiencing, have the same type of abstract
characterization and the same kind of physical structure and, because of
the ``no collapse'' hypothesis, all those physical structures are ``real'' parts
of the universal state.  On the basis of these similarities alone, I would be
inclined to accept the actuality both of all of your minds and of all of
mine.

As Lockwood 1996a points out, a theory in which only one mind is actual
for each individual would be a hidden variable theory, with actuality as
the hidden variable.  In such a theory, if, in the usual way, I measure an
up-spin in one electron from a singlet state in a EPR-Bohm situation, then,
when you, at a spacelike separation, measure the spin of the other
electron, either all the well-known non-locality problems arise in forcing
the ``actual'' you to see a down-spin, or I can encounter a ``non-actual''
you -- a ``mindless hulk'' (Albert 1992).  This is bad enough, but such a
theory would also require the identification of the set of ``individuals''.
At a technical level, I do not know how this could be done.  In a theory
with a ``consciousness'' basis, how am I to decide to whom a given
element of the basis belongs?  In an appropriate set of $10^{10^{26}}$
orthogonal wavefunctions available to a warm wet brain, there is an
imperceptible passage from wavefunctions which represent me to those
which represent you.  Where is this line to be drawn?  Even in my theory,
in which individuals have pasts, all those pasts are ultimately
undifferentiated.  And finally, we cannot identify the set of all individuals
present at a given moment, unless we are prepared to define ``a given
moment''.  To do that would be to attempt a ``many-worlds'' theory not a
``many-minds'' theory.  At the very least, incompatibility with special
relativity would surely follow.

\medskip
\centerline{\bf Identity over time}
\medskip

	  There has also been much discussion in the literature of the problem of
identity over time  (e.g. Butterfield 1996, \S 4).  In this connection, I
note that at the heart of my technical construction there is a definition
(Donald 1990, Hypothesis V and 1995, Definition E), which attempts to
capture, within the mathematics of quantum field theory, the idea of an
object existing through time by looking for the paths in spacetime along
which the local quantum state changes least.  This is a very direct
approach to the problem, and one which shows that identity over time
can be defined without even using the idea of an ``object'' as being
something composed of particles.

Unfortunately, there is a technical flaw in the application of this definition
in my 1995 paper.  The problem is that I have not allowed for the effect,
on the path under consideration, of information gained elsewhere in the
brain.  I believe, however, that I can solve this problem fairly
straightforwardly by disconnecting the times of quantum state change
from the times of object ``switching''.  Work is in progress on the details
of this revision and I hope in due course to publish it, together with the
details of the two other suggestions I have described above; one, to allow
the experience of an individual observer to be modelled as the experience
of observing a particular, identified, discrete, stochastic process, and the
other, to provide a physics without physical constants.

\bigskip

\noindent{\bf Acknowledgements } I am grateful to Jeff Barrett,
Katherine Brading, Jeremy Butterfield, Michael Esfeld, Michael Lockwood,
Simon Saunders, and Steve Weinstein for useful conversations and
comments.

\bigskip
\centerline{\bf References}
\medskip
\frenchspacing
\parindent=0pt

\rf Albert D.: 1992,  {\sl Quantum Mechanics and Experience,} Harvard
University Press.

\rf Albert, D and Loewer, B.: 1988, `Interpreting the Many Worlds
Interpretation', {\sl Synthese, \bf 77,}  195--213.

\rf Bacciagaluppi, G., Donald, M.J., and Vermaas, P.E.: 1995, `Continuity
and Discontinuity of Definite Properties in the Modal Interpretation', {\sl
Helvetica Physica Acta, \bf 68,}  679--704.

\rf Barrett, J.A.: 1995, `The Single-Mind and Many-Minds Versions of
Quantum Mechanics', {\sl Erkenntnis, \bf 42,}  89--105.

\rf  Brown, H.R.: 1996, `Mindful of Quantum Possibilities',
{\sl  British Journal for the Philosophy of Science, \bf 47,}  189-200.

\rf Butterfield, J.: 1995, `Worlds, Minds, and Quanta', {\sl Aristotelian
Society Supplementary Volume, \bf 69,}  113--58.

\rf Butterfield, J.: 1996, `Whither the Minds?',
{\sl  British Journal for the Philosophy of Science, \bf 47,}  200--21.

\rf Churchland, P.S. and Sejnowski, T.J.: 1992,  {\sl  The Computational
Brain,} Bradford,  Cambridge MA.

\rf  Dennett, D.C.: 1991,  {\sl  Consciousness Explained,} Little
Brown,  Boston.

\rf Deutsch, D.: 1996, `Comment on Lockwood',
{\sl  British Journal for the Philosophy of Science, \bf 47,}  222--8.

\rf  DeWitt, B.S. and Graham, N.: 1973,  {\sl  The Many-Worlds
Interpretation of Quantum Mechanics,} Princeton University Press.

\rf  Dickens, C.: 1843,  {\sl  A Christmas Carol,} Chapman and Hall,
London.

\rf   Donald, M.J.: 1990, `Quantum Theory and the Brain',  {\sl
Proceedings of the Royal Society of London, \bf A 427,} 	43--93.

\rf  Donald, M.J.: 1992, `A Priori Probability and Localized Observers',
{\sl   Foundations of Physics, \bf 22,} 	1111--72.

\rf  Donald, M.J.: 1995, `A Mathematical Characterization of the Physical
Structure of Observers', {\sl   Foundations of Physics, \bf 25,}
529--71.

\rf  Dowker, F. and Kent, A.: 1996, `On the Consistent Histories Approach
to Quantum Mechanics', {\sl   Journal of Statistical Physics,
\bf 82,}  1575-1646.

\rf Gell-Mann, M. and Hartle, J.B.: 1993, `Classical Equations for
Quantum Systems', {\sl Physical Review, \bf D 47,}  3345--82.

\rf Haag, R.: 1992,  {\sl Local Quantum Physics,} Springer-Verlag, Berlin.

\rf  Kent, A.: 1990, `Against Many-Worlds Interpretations', {\sl
International Journal of Modern Physics,  \bf A 5,}  1745-62.

\rf Lehner, C.: 1997,  `What It Feels Like to be in a Superposition and
Why',  {\sl Synthese, \bf 110,}  191--216.

\rf  Lewis, D.: 1986,  {\sl  On the Plurality of Worlds,} Blackwell,
Oxford.

\rf  Lockwood, M.: 1989,  {\sl  Mind, Brain, and the Quantum,} Blackwell,
Oxford.

\rf  Lockwood, M.: 1996a, ``Many Minds' Interpretations of Quantum
Mechanics', {\sl  British Journal for the Philosophy of Science, \bf 47,}
159--88.

\rf  Lockwood, M.: 1996b, ``Many Minds' Interpretations of Quantum
Mechanics: Replies to Replies', {\sl  British Journal for the Philosophy of
Science, \bf 47,}  445--61.

\rf  Loewer, B.: 1996, `Comment on Lockwood', {\sl  British Journal for
the Philosophy of Science, \bf 47,}  229--32.

\rf  Papineau, D.: 1995,  `Probabilities and the Many Minds Interpretation
of Quantum Mechanics', {\sl   Analysis, \bf 55,}  239--46.

\rf  Papineau, D.: 1996, `Many Minds are No Worse than One', {\sl  British
Journal for the Philosophy of Science, \bf 47,}  233--41.

\rf  Page, D.N.: 1996, `Sensible Quantum Mechanics: Are Probabilities Only
in the Mind?', {\sl  International Journal of Modern Physics, \bf D
5,}  583--96.

\rf  Saunders, S.: 1993, `Decoherence, Relative States, and Evolutionary
Adaptation', {\sl   Foundations of Physics, \bf 23,}  1553--85.

\rf  Saunders, S.: 1995, `Time, Quantum Mechanics, and Decoherence',
{\sl   Synthese, \bf 102,}  235--66.

\rf  Saunders, S.: 1996a, `Comment on Lockwood', {\sl  British Journal
for the Philosophy of Science, \bf 47,}  241--8.

\rf  Saunders, S.: 1996b, `Relativism',  125--42, in R. Clifton {\sl (ed),
Perspectives on Quantum Reality,}  Kluwer, Dordrecht.

\rf  Silverman, M.P.: 1995,  {\sl  More than One Mystery: Explorations in
Quantum Inference,} Springer-Verlag, Berlin.

\rf Tonomura, A., Endo, J., Matsuda, T., and Kawasaki, T.: 1989,
`Dem\-onstration of Single-Electron Build-Up of an Interference Pattern',
{\sl Amer. J. Phys., \bf 57,}  117--20.

\rf Weinstein, S.: 1996,  `Undermind',  {\sl Synthese, \bf 106,}
241--51.

\end